\documentclass[]{article}
\usepackage{multicol} 
\usepackage[utf8]{inputenc}
\usepackage{graphicx}
\usepackage{booktabs} 
\usepackage{float} 
\usepackage{bm} 
\usepackage[hmarginratio=1:1,top=32mm,columnsep=20pt]{geometry} 
\usepackage{abstract} 
\usepackage{titlesec} 
\usepackage{upgreek}
\usepackage{color}
\usepackage{hyperref} 
\usepackage{amsmath,amssymb} 
\usepackage[caption=false]{subfig}
\usepackage{units}
\usepackage{authblk}
\usepackage{color}
	\hypersetup{
  final,
  pdfborder={0 0 0},
  pdftitle={Deformation induced complexion transitions  nanocrystalline  alloys},
  pdfauthor={Deckarm, Sch"afer, Albe, Birringer},
  colorlinks=true,
  urlcolor=red,
  linkcolor=black,
  citecolor=blue
}

\begin{document}
\title{Deformation induced complexion transitions  in nanocrystalline  alloys}
\author[1]{Michael Johannes Deckarm\thanks{michael.deckarm@nano.uni-saarland.de}}
\author[2]{Jonathan Sch\"afer\thanks{schaefer@mm.tu-darmstadt.de}}
\author[2]{Karsten Albe\thanks{albe@mm.tu-darmstadt.de}}
\author[1]{Rainer Birringer\thanks{rainer.birringer@nano.uni-saarland.de}}

\affil[1]{\small{Universit\"at des Saarlandes, FR7.2 Experimentalphysik, Campus D2 2, D-66123 Saarbr\"ucken}}
\affil[2]{\small{Technische Universit{\"a}t Darmstadt, Fachbereich Material- und Geowissenschaften, Fachgebiet Materialmodellierung, Jovanka-Bontschits-Str. 2, D-64287 Darmstadt}}


\maketitle

\begin{abstract}
\noindent Grain boundary (GB) enthalpies in nanocrystalline (NC) $\mathrm {Pd_{90}Au_{10}}$ are studied after preparation, thermal relaxation and plastic deformation.
By comparing results from  atomistic computer  simulations and calorimetry, we show  
that increasing plastic deformation of equilibrated NC $\mathrm {Pd_{90}Au_{10}}$ specimen causes 
an increase of the stored GB enthalpy $\Delta \gamma$. We interpret this change of $\Delta \gamma$ as stress-induced complexion transition from a low-energy to a high-energy GB-core state. In fact, GBs behave not only as mere sinks and sources of zero- and one-dimensional defects or act as migration barriers to the latter but also have the capability of storing deformation history through configurational changes of their core structure and hence GB enthalpy. Such a scenario can be understood as a continuous complexion transition under non-equilibrium conditions, which is related to hysteresis effects under loading-unloading conditions.
\end{abstract}


\begin{multicols}{2} 



Over the past three decades, it has been well established that decreasing the grain size of polycrystalline metals into the nanometer regime results in a substantial increase of strength that has eventually led to coin the slogan "smaller is stronger" \cite{Meyers2006}. For average grain sizes at the lower end of the nanoscale $<$ 10 nm, it has been found that intragranular crystal plasticity becomes to a large proportion replaced by intergranular plasticity, where deformation processes essentially emerge in the core region of GBs. 
Computer simulations and experiments unraveled a variety of modes of plastic deformation that are mediated by GBs: GB slip and sliding \cite{VoAverBel08, VanSwygenhoven2001, Weissmuller2011}, stress-driven GB migration \cite{Cahn2004a, CahMisSuz06, SchaAlb12}, grain rotation \cite{Legros2008} as well as shear transformations (STs) localized in the core regions of GBs \cite{Lund2005, Argon2006}. The latter involve shuffling or flipping of groups of atoms and may act as flow defect in the core region of GBs, thus playing a similar role in a disordered proximity as dislocations do in a crystalline environment. 


While in conventional crystal plasticity texture formation is an indicator for deformation induced intracrystalline dislocation activity, in the context of NC metals the fundamental 
question  arises, whether or not GBs also entail the capability of storing deformation history through irreversible configurational changes of the GB-core structure and hence GB enthalpy. Recently, it has been shown that changes of GB configurations can be universally described by the concept of complexions \cite{Cantwell20141}, where the GB enthalpy serves as an order parameter. A complexion represents an equilibrium state or a steady-state situation of matter at a crystalline interface which is neither amorphous nor crystalline. 
In analogy to phase transitions in the bulk, a complexion transition is defined as a discontinuity  in the first or higher order derivatives of the  interface-energy functional with respect to the thermodynamic parameters temperature, pressure, chemical potentials and crystallographic orientation parameters.
Temperature induced transitions and hysteresis effects have been reported for GB states in various materials 
\cite{Cantwell20141}. In a recent study utilizing MD simulations, Frolov et. al. \cite{Frolov2013} have discovered, at temperatures well below the melting point, a congruent transition in high angle GBs in pure Cu where the atomic structure of the grain boundary core changed but the GB geometry remained unchanged. Likewise, recent experimental measurements of Ag impurity diffusion along GBs in Cu revealed a discontinuity in the activation energies (non-Arrhenius behavior) \cite{Divinski2012}. Atomistic simulations showed that this is due to a structural phase transformation in the core region of high-angle GBs \cite{Frolov2013a}. We are not aware, however, of any study dealing with stress-induced complexion transitions. These should favorably occur in nanoscale materials in the limit of small grain sizes, where GB-mediated deformation mechanisms prevail \cite{Grewer2014}. 
While a recent simulation study implies that such transitions do exist \cite{SchStuAlb11}, it is a priori not obvious, if they are continuous or discontinuous by nature and experimentally
detectable.

Dealing with stress-induced complexion transitions, the aim of this study is to compare results from atomistic computer simulations and experiment on a quantitative basis. Our atomic scale computer simulations provide full access to structural and related energetic changes and at the same time to macroscopically observable quantities which are amenable to experimental validation; here we concentrate on stored GB enthalpy. Combining simulation and experiment enables to discriminate processes related to changes of the energetics of the core structure of GBs from degrees of freedom of storing or annihilating energy  associated with the crystalline phase or migrating GBs. To furnish evidence of a complexion transition induced by stress (pressure) necessitates verification of a distinct change of GB enthalpy upon plastic deformation.

%
\begin{figure*}[htpb]
\includegraphics[width=1.05\textwidth]{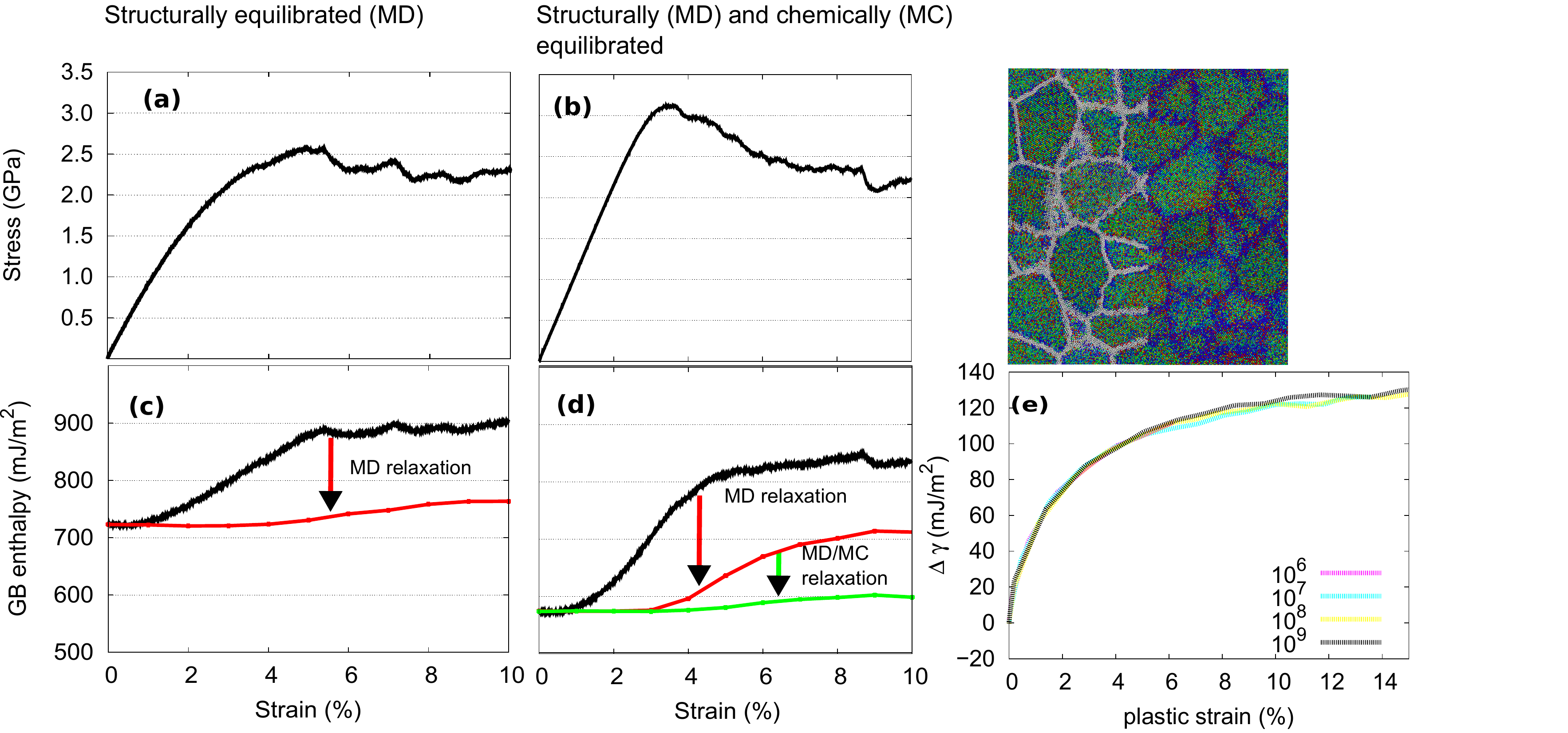} 
\caption{Stress-strain behavior of samples with different initial grain boundary (GB) states: Panel (a) random solute distribution, structurally equilibrated by MD before deformation; Panel (b) structurally (MD) and chemically (MC) equilibrated GBs before deformation. Panels (c) and (d): to (a) and (b) related excess enthalpies normalized by the initial GB area. Black curve: excess enthalpy during straining, red curve: excess enthalpy after relaxing the structure to zero stress, green curve in (d): excess enthalpy after equilibrating the relaxed structure via MD/MC. Subtracting the green from the red curve yields the stress-dependent change of GB enthalpy $\Delta \gamma$ shown in panel (e) for different strain rates. The snapshot (top right) represents a slice through the initial microstructure (d=10~nm). Red denotes enrichment and blue depletion with respect to global composition (10\%~Au)). GB atoms are highlighted in white on the left half of the slice. Snapshots were generated using OVITO. \cite{Stu10} }
\label{fig:str_enthalpy}
\end{figure*}


%

In this study, we investigate the evolution of the average grain boundary enthalpy in NC PdAu, which is a fully miscible alloy system, upon mechanical loading. Precious metals are used to ensure that impurity effects and energy contributions from possible chemical reactions (e.g. formation of precipitates) play -- if at all -- a minor role with regard to stress-induced changes of the energetics of GBs. In fact, it is essential to study the NC $\mathrm {Pd_{90}Au_{10}}$ alloy system, where solute drag of Au renders the NC material stable against curvature-driven grain growth up to $\unit[130]{^{\circ} C}$; pure NC Pd is configurationally unstable and even exhibits pronounced grain growth at room temperature \cite{Ames2008}. 

We  begin with reporting the results from computer simulations.  
Model structures consisting of 54 grains with 10 nm average grain size were compressed at \unit[300]{K} by molecular dynamics simulations.
We imposed a constant engineering strain rate ($10^8$~1/s) in uniaxial direction on the simulation cell. The cell size was allowed to relax perpendicular to the strain axis.
For relaxing the model system before and after deformation, we applied a hybrid simulation method 
that accounts for structural relaxations and thermal vibrations as well as exchange of atom types \cite{SadErhStu12}.
All details on the method, the interatomic potentials and structure generation are given in Ref. \cite{SchStuAlb11} and in the supplement.



The stress-strain behavior of a structurally (MD) equlibrated sample is displayed in Fig. \ref{fig:str_enthalpy} a) and an additionally chemically (MD/MC) equilibrated sample is shown in Fig. \ref{fig:str_enthalpy} b). The changes of specific (per unit area of GB) enthalpy  (black lines) related to irreversible plastic deformation are displayed in Fig. \ref{fig:str_enthalpy} c) and d), respectively. Comparison of the two sample states clearly reveals a comparatively higher specific enthalpy ($\approx \unit[700]{mJ/m^2}$) of the only structurally equilibrated sample in the undeformed state. The fully (structurally and chemically) equilibrated samples assume lower values of stored GB enthalpy ($\approx \unit[600]{mJ/m^2}$) originating from rearrangements of the solutes in the GB core region. Upon loading, the specific enthalpy increases more steeply for the fully equilibrated sample in agreement with a steeper increase of applied stress and a significantly higher yield stress \cite{SchStuAlb11}.

The evolution of specific enthalpy as a function of applied strain results from different contributions: the elastic energy due to the applied load, energy contributions of lattice defects and possible changes of the GB enthalpy.
The elastic energy can simply be released by removing the external load and relaxing the structure to zero hydrostatic stress within a standard MD run. The remaining specific enthalpies (red lines in Fig.~\ref{fig:str_enthalpy} c) and d) ) contain energy contributions of irreversible GB transitions  and of residual lattice defects. They show a more pronounced increase with plastic strain for the fully equilibrated sample as compared to the only structurally equilibrated one. In any case, this observation implies that accommodation of plastic strain, involving transient dilatancy and shear shuffling in GBs, sensitively depends on the configurational state of the core region of GBs. 

In order to extract the enthalpy contributions due to GB transitions, the initially fully equilibrated sample was again chemically and structurally re-equilibrated. Since existent lattice defects cannot be annealed within the given time scales at a temperature of 300 K, any change in potential energy must then result from chemically and structurally equilibrating the GBs. Hence, we assign the difference between the red and green curves in  Fig.~\ref{fig:str_enthalpy} d) to the change in GB enthalpy $\Delta \gamma$ due to configurational changes (transitions) in the core region of GBs which have been induced by the external loading conditions. The slight increase in specific enthalpy of the fully equilibrated sample with strain (green line)  originates from residual lattice defects. In Fig.~\ref{fig:str_enthalpy} e), we display the dependence of $\Delta \gamma$ on $\varepsilon_{\mathrm{plastic}}$. For all tested strain rates, we observe the same functional behavior with a tendency to stagnation at large plastic strains.




\begin{figure}[H]
\centering
\includegraphics[width=\columnwidth]{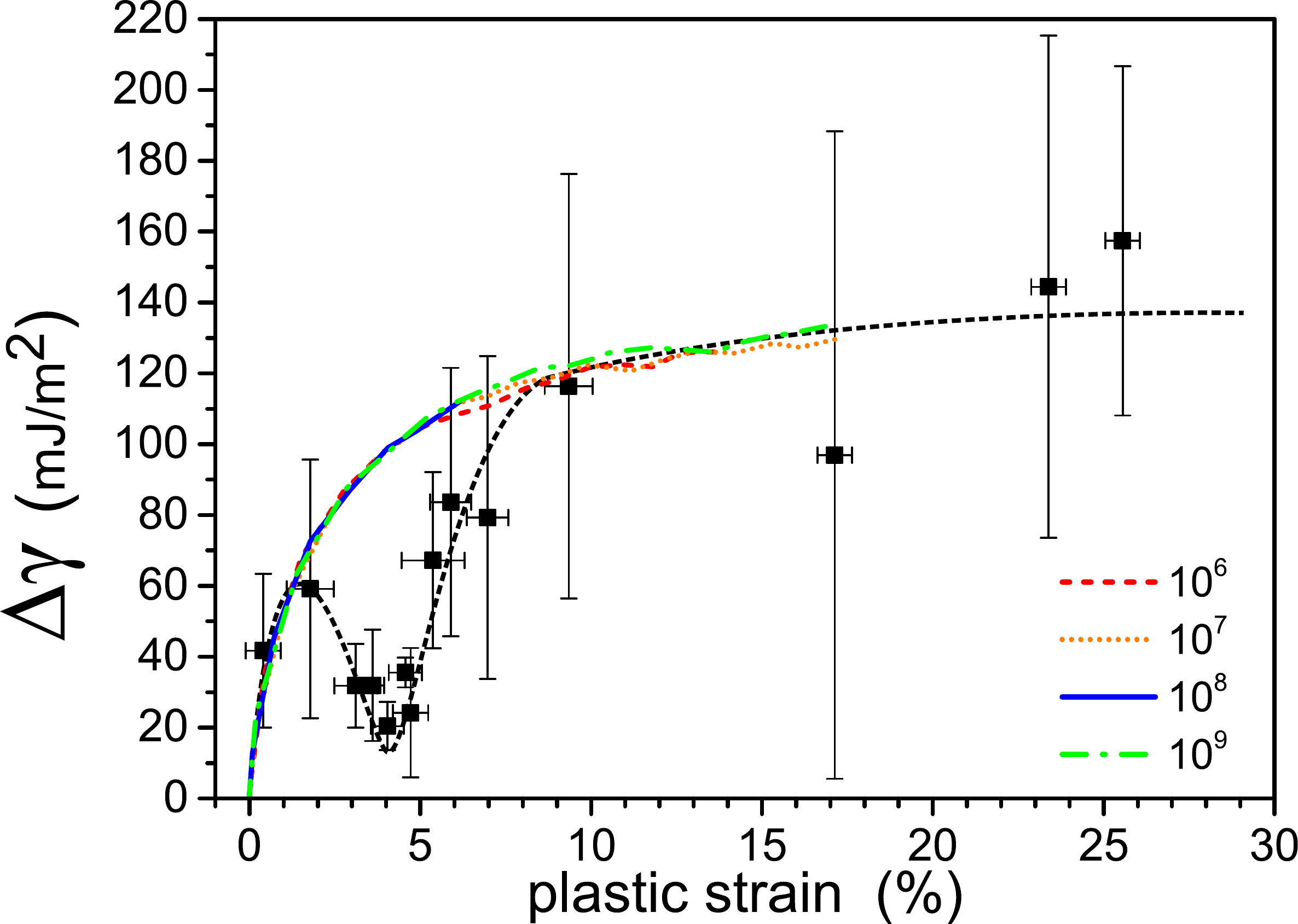}
\caption{Change of GB enthalpy $\Delta \gamma$ of relaxed and differently intense deformed samples vs. plastic strain. Full black squares represent experimental data (calorimetry) and colored lines results from simulations (Fig. \ref{fig:str_enthalpy} e). The black dashed line serves as a guide to the eye for the experimental data.}
  \label{fig:plstr_change}
\end{figure}

To validate the central result (Fig.~\ref{fig:str_enthalpy} e)) and reasoning deduced from the computer simulation studies, we  deformed a set of thermally equilibrated (relaxed) NC ${\rm Pd_{90}Au_{10}}$ samples with a grain size of 
$\approx \unit[10]{nm}$ by applying uniaxial compression. After unloading different specimens of the set at gradually increasing plastic strain, released enthalpies have been extracted by means of calorimetry. In parallel, we employ X-ray diffraction enabling to deduce the associated GB-area. We are aware that relating the exothermic heat flux originating from irreversibly strained samples to stored enthalpy in GBs implies that heat flux contributions from the interior of the nanograins should be negligibly small, which will later be justified. A priori, we can not rule out stress-driven grain boundary migration \cite{CahMisSuz06} as a possible strain carrying deformation mode \cite{Grewer2014}. Therefore, we carefully analyze the grain size that may change due to loading. The variation of enthalpy $\Delta\gamma$, that has been induced through applying plastic strain, is taken relative to the relaxed but undeformed specimen state, serving as reference state. We project the released enthalpy $\Delta H$ onto the related interface area measured after unloading, $A_{\mathrm{GB}}$, to then obtain the GB enthalpy $\Delta  \gamma $ from the relation $\Delta \gamma = {\Delta H} / {A_{\text{GB}}}$. Details related to sample preparation and methodology (calorimetry and XRD) are reported in the supplement.

\begin{figure}[H]
\centering
\includegraphics[width=\columnwidth]{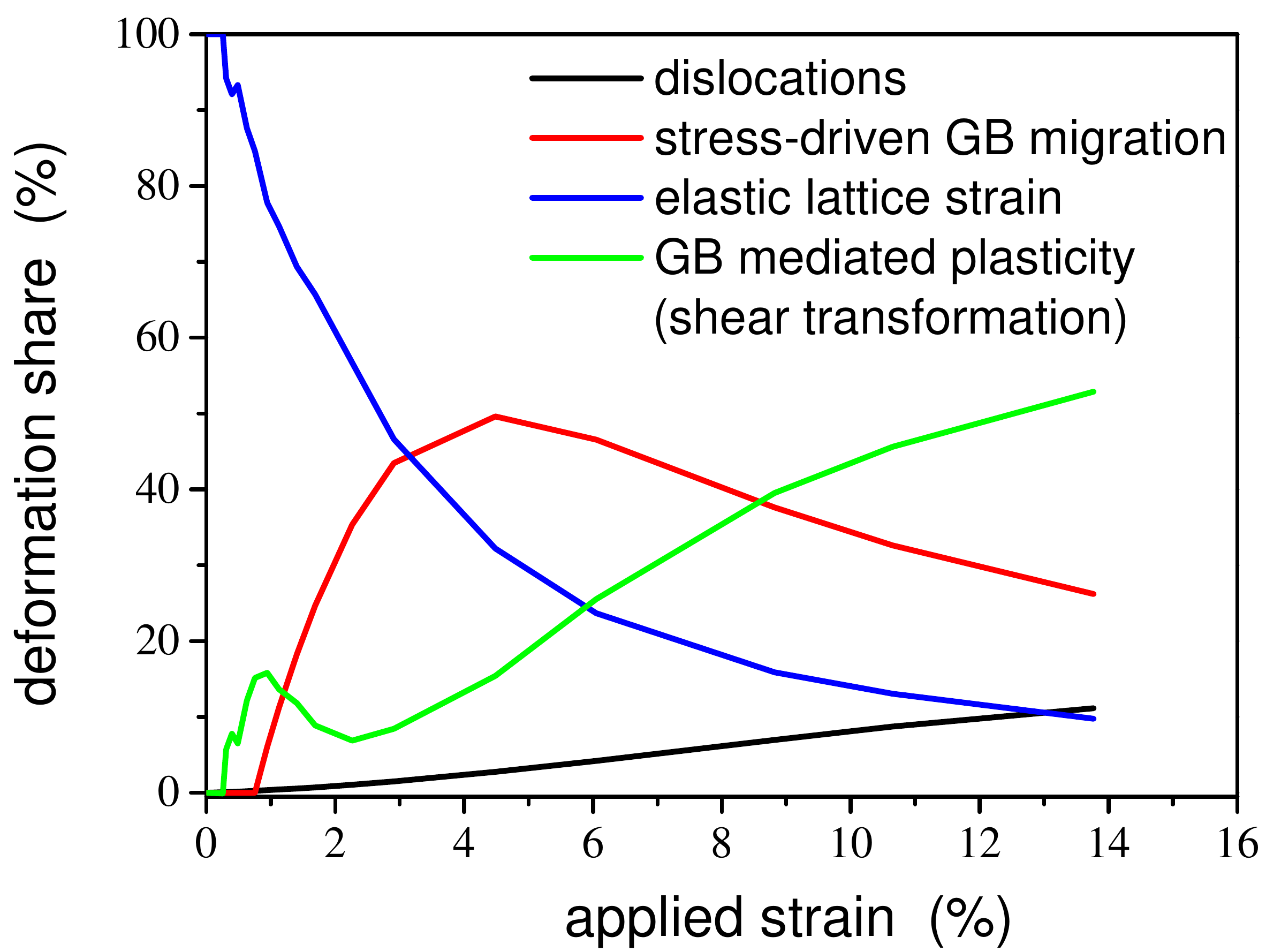}
\caption{Deformation share of different strain carrying mechanisms vs. applied strain; shares are specified relative to the applied strain. More detailed information on used methods and data analysis is given in \cite{Grewer2013, Grewer2014}.} 
  \label{fig:manuel}
\end{figure}


The dependence of $\Delta \gamma$ on $\varepsilon_{\mathrm{plastic}}$ is shown in Fig. \ref{fig:plstr_change}. Except for the cusp at $\varepsilon_{\mathrm{plastic}}= 4\%$, the deformation induced increase of the GB enthalpy
exhibits a functional behavior with a tendency to stagnation at the largest strain values. Comparing the deformation induced enthalpy changes with the simulated data discussed before, we find, apart from the cusp, a strikingly good agreement between the computer simulations and the experimental data. In both cases, plastic deformation results in a cumulative storage of enthalpy in the GBs. The existence of the cusp in the experimental results seems to contravene this finding but we are going to argue that this is in fact just an apparent contradiction. 

The onset of cusp formation ($\varepsilon_{\mathrm{plastic}} \gtrsim 2\%$) coincides with stress-driven GB migration (SDGBM), which is according to Grewer \cite{Grewer2013} the dominant deformation mechanism in this strain regime (see Fig. \ref{fig:manuel}). Since migrating GBs most likely have assumed local equilibrium after unloading, they exist in the equilibrated reference state and thus do not contribute to exothermic heat flux. Consequently, they generate no share to the ensuing enthalpy measurements, as a result $\Delta \gamma$ declines and so opens up the cusp. In view of this behavior, we presume that GBs which have been energized at low strain most effectively are those which preferably contribute to SDGBM. Obviously, in addition to thermal relaxation of GBs, SDGBM establishes still another scenario for mechanical relaxation/equilibration of GBs.

For strain values $\varepsilon_{\mathrm{plastic}} \gtrsim 4\%$, SDGBM ceases, which manifests through a continuous decrease of the deformation share of SDGBM as shown in Fig. \ref{fig:manuel}. The concomitant cessation of mechanical equilibration of GBs results again in a strain-induced gradual increase of GB enthalpy. 
Since SDGBM is not observed in the simulated sample, the cusp in $\Delta \gamma$ does not occur and the simulation reveals continuous storing of GB enthalpy.
 
The minimum of the cusp yields yet another essential information regarding additional sources of exothermic heat flux. Migrating GBs can sweep-off lattice defects, particularly dislocations or stacking faults \cite{Skrotzki20137271}, thereby causing an additional exothermic heat flux which would however counteract cusp formation. Therefore, the cusp is not only evidence that dislocations and stacking faults play a minor role in the calorimetrically observable enthalpy changes in NC $\mathrm {Pd_{90}Au_{10}}$ \cite{Skrotzki20137271}, it unambiguously underlines the dominant influence of the GB structural state on $\Delta \gamma$.

%
The central result of this study (Fig. \ref{fig:plstr_change}) clearly reveals that increasing plastic deformation of equilibrated NC $\mathrm {Pd_{90}Au_{10}}$ specimens causes an increase of the stored GB enthalpy $\Delta \gamma$. The non-abrupt increase of $\Delta \gamma$ implies that the ensemble of GBs undergoes a continuous transition from a low-energy to a high-energy core state. The latter configuration prevails in the macroplastic regime of deformation. Here, the GB enthalpy, which acts as an order parameter,
is stagnating to a constant value.

 Overall, this behavior can be understood as a continuous complexion transition under non-equilibrium conditions, which is related to hysteresis effects under loading-unloading conditions \cite{Schaefer2013}. We are not aware of any prediction from theory concerning the scaling behavior of $\Delta \gamma$ versus stress (pressure) or strain, respectively. 

For the sake of completeness, we notice that the variation of grain boundary enthalpy scales as  $(\Delta \gamma \propto 1 -\exp{(-\varepsilon_{\mathrm{plastic}})})$ 
. Clearly, the low-energy core state represents GBs in local equilibrium (reference state). Their specific grain boundary enthalpy $\gamma_{\mathrm{low}} \simeq \unit[600]{mJ/m^2}$ compares favorably with values which are characteristic of equilibrium high angle GBs \cite{SutBal94}. The high-energy core state $\gamma_{\mathrm{high}} \gtrsim \unit[750]{mJ/m^2}$ appearing in the regime of stagnation must be correlated with configurational changes caused by GB-mediated deformation processes. As seen in Fig.~\ref{fig:manuel}, deformation is dominated by strain propagation through shear shuffling at/along GBs \cite{Grewer2014} in this regime. The generic flow defect enabling GB-mediated deformation are localized shear transformations involving 20-30 atoms which generate incremental strain associated with GB excess-volume fluctuations \cite{Grewer2014, Argon1979}. Mechanistically, the simultaneous presence of GB-shear and excess volume fluctuations are prerequisites for the emergence of structural transitions. We therefore conjecture that the transition from the $\gamma_{\mathrm{low}}$ to the $\gamma_{\mathrm{high}}$ state is related to a transition of structural units from their ground state into novel configurations. 
 
The observed stagnation indicates that propagation of strain is related to a saturation of excess volume fluctuations in GBs \cite{SchStuAlb11}. This state minimizes the stress threshold for crossing the activation barrier of shear transformations. Since GB cores are characterized by atomic-site mismatch and consequently short-range disorder, we note that the raised issue also has a link to stress-induced rejuvenation observed in metallic glasses  \cite{Meng2012}. 

Finally, we conclude that GBs behave not only as mere sinks and sources of zero- and one-dimensional defects or act as migration barriers to the latter but also have the capability of storing deformation history through configurational changes of their core structure and hence GB energy. The stress-induced continuous transition from a low-energy to a high-energy GB-core state can be perceived as a GB complexion transition.

The authors acknowledge the financial support of the Deutsche Forschungsgemeinschaft (FOR714 and 385/18-1) and the grants of computer time from John von Neumann Institute for Computing in J\"ulich. We thank K.-P. Schmitt (INM – Leibniz-Institut für Neue Materialien gGmbH, Campus D2 2, 66123 Saarbrücken, Germany) for the help and support in carrying out the deformation experiments.

\newpage

\section{Methodology: Simulations}

\subsection{Interatomic potential}
\label{sec:potential}

For studying the Pd--Au alloy at the atomic level, we employ a combination of molecular dynamics (MD) and Monte-Carlo (MC) methods based on semi-empirical interatomic potentials. Pure elements are described by EAM potentials for Pd \cite{FoiHoy01} and Au \cite{AckTicVit87}. The cross interaction follows a recent formulation of the \emph{concentration-dependent embedded-atom method} (CD-EAM) \cite{StuSadErh09,CarCarLop06} that has been developed to exactly reproduce the enthalpy of mixing of alloys over the full composition range. 
To make the EAM potentials for the pure elements compatible and to minimize nonlinear contributions of the embedding terms to the formation energy, we first normalized the Pd and Au EAM potentials to an \emph{effective pair representation} \cite{CarCarLop06} that preserves all properties of the pure elements. The five coefficients of the $h(x)$ polynomial were then determined such that the experimental enthalpy of mixing curve at \unit[1300]{K} \cite{LandoltBornsteinPdAu} is reproduced by the potential for a random solid solution. Table~\ref{tab:HCoeff} lists the coefficients of the $h(x)$ polynomial after fitting.
\begin{table}[H]
	\caption{Coefficients of the 4th order polynomial $h(x)=\sum^{4}_{n=0}{h_n x^n}$ for the Au--Pd potential. Here, $x$ denotes the local Au concentration ($0 \leq x \leq 1$).}
	\begin{tabular}[c]{cc}
		\\
		$h_0$ & $\phantom{-}1.159085$ \\
		$h_1$ & $-0.126781$ \\
		$h_2$ & $\phantom{-}0.481763$ \\
		$h_3$ & $-0.488693$ \\
		$h_4$ & $\phantom{-}0.203778$
	\end{tabular}
	\label{tab:HCoeff}
\end{table}

\subsection{Preparation of nanocrystalline model structures}
\label{sec:preparation}

We first created nanocrystalline model structures with an average grain size of 10 \unit{nm} consisting of 128 grains, respectively. The Voronoi method \cite{Vor08} was used to set up the grain shapes based on randomly placed center points in a cubic simulation box. Three-dimensional periodic boundary conditions were applied to model a bulk structure without free surfaces. The lattice orientations of the grains were drawn from a random isotropic distribution. To avoid spurious high-energy atomic configurations in the as-prepared Voronoi samples, we deleted atoms from the grain boundaries that were closer than \unit[2.0]{\AA} to other atoms prior to relaxation.

To distinguish atoms form the grain boundaries from normal bulk atoms, we apply the common neighbor analysis (CNA) \cite{HonAnd87} method. The cutoff parameter $R_{\textrm{CNA}}$ for nearest neighbors was set between the first and second fcc neighbor shells: $R_{\textrm{CNA}}=a_0(x) \cdot (1+\sqrt{1/2})/2$, with $a_0(x)$ being the \unit[0]{K} lattice parameter for a given concentration $x$.

\subsection{Mechanical testing}

We deformed all samples at \unit[300]{K} by imposing a constant engineering strain rate ($10^6$ to $10^9$~1/s) in uniaxial direction on the simulation cell. The cell size was allowed to relax perpendicular to the strain axis.

\subsection{MD/MC relaxation}
\label{sec:alloying}

For relaxing the structure before and after deformation, we utilize a hybrid simulation method that accounts for structural relaxations and thermal vibrations as well as chemical mixing and ordering. One part of this technique is a transmutational Monte-Carlo scheme that samples the semi-grandcanonical ensemble, and thus allows us to determine the equilibrium concentration and element distribution in the sample for a given  chemical potential difference $\Delta \mu=\mu_{Au}-\mu_{Pd}$. At the same time, we account for structural relaxations in our polycrystalline setting by interleaving molecular dynamics (MD) steps: After the system has been evolved for a certain number of MD steps, we carry out MC trial moves, before returning to the MD part again. The number of MD steps between MC moves as well as the number of MC trial moves can be adjusted to optimize the convergence of the simulation. Using this combination of simulation methods, vibrational effects on the mixing behavior are naturally included in the model.

We use the freely-available molecular dynamics code LAMMPS \cite{Pli94}, which we have extended to perform a simple transmutation Monte-Carlo algorithm: Atoms picked at random are replaced by atoms from the other species and the resulting energy change is calculated. Such transmutation trial moves are accepted or rejected based on the Metropolis algorithm, with a transition matrix that samples the semi-grandcanonical ensemble. In case of an accepted type swap, the velocity of an atom is rescaled to conserve its kinetic energy. Both, the MD and MC parts of the simulation code are parallelized to enable large-scale simulations of reasonably sized nanocrystalline alloy structures.

Relaxation and alloying was performed at \unit[300]{K} for \unit[1]{ns} at zero hydrostatic pressure using Berendsen's \cite{BerPosGun84} thermostat and barostat. During this MD run, one full MC step was performed every \unit[40]{fs}, i.e., 25{,}000 trial exchanges were performed on each atom in the system on average. The temperature parameter for the Metropolis MC algorithm was also set to \unit[300]{K} as in the MD stage. The distribution of solute was therefore fully equilibrated within the microstructure. This sample will therefore be referred to as equilibrated by MD/MC throughout the manuscript.
To serve as a reference, an additional sample was prepared, were the same amount of solute atoms was introduced at random sites throughout the microstructure. The sample was then equilibrated by MD only at \unit[300]{K} and zero hydrostatic pressure. This reference sample will be referred to as equilibrated by MD throughout the manuscript.

\section{Methodology: Experiment}

\subsection{Sample Preparation}

The NC PdAu-alloy specimens with approximately \unit[10]{at\%} Au  (${\rm Pd_{90}Au_{10}}$) were prepared by inert gas condensation (IGC) and compaction \cite{Birringer1984}. Compaction of the assembled nanocrystals took place in an piston and anvil device under high vacuum ($< \unit[10^{-6}]{mbar}$) and a compaction pressure of \unit[1.85]{GPa} during a time period of \unit[30]{s}. The composition of the so received specimens was determined by EDX in a SEM (JEOL 7000F) to find a deviation of $\pm 2\%$ from the nominal $\mathrm {Pd_{90}Au_{10}}$ composition. Their geometry is disc-shaped with a diameter of \unit[8]{mm} and a thickness of about \unit[1]{mm}. These disc-shaped pellets are then cut into 4 squares with \unit[2]{mm} edge length and are referred to as "as-prepared" samples.

\subsection{Mechanical deformation}
Mechanical deformation was performed as uniaxial compression test in a Zwick Typ 1476 mechanical materials testing machine with a strain rate of $10^{-4}$ ~1/s. Stepwise increasing strain was applied individually on each sample to reach the intended plastic strain value. Plastic strain was determined with a micrometer caliper after each step.

\subsection{X-ray diffraction}

Grain size was determined by XRD measurements using a Panalytical X'Pert Pro diffractometer with Cu-cathode and PIXcel-1D detector in $\theta - \theta$ geometry. X-ray spectra covering a $2 \theta$ range from $30^{\circ}$ to $138^{\circ}$ were recorded with $0.026^{\circ}$ step size and $\unit[400]{s}$ measuring time per step. The mean grain size $\langle D \rangle_{\mathrm{vol}}$ and micro strain were determined by the Klug and Alexander method as described in \cite{Klug1974}. In the course of this work, we need to determine the overall GB area, $A_{\mathrm{GB}}$. For the ratio grain boundary area per unit volume of crystal $A_{\mathrm{GB}} / V$, stereology delivers the following identity: $A_{\mathrm{GB}} / V = 2 / \langle L \rangle_{\mathrm{area}} $, where $\langle L \rangle_{\mathrm{area}}$ is the area-weighted average column length of grains, which is just a specific moment of the grain size distribution function. Following Krill et al. \cite{Krill1998}, $\langle L \rangle_{\mathrm{area}}$ can be converted into the measured volume-weighted average grain diameter to obtain ${A_{\mathrm{GB}}} = m/\rho \cdot {3}/{\langle D \rangle_{\mathrm{vol}}} \exp \left( - \ln^2 \sigma \right)$, where $m$ is the sample mass, $\rho$ its density, and $\sigma \approx 1.7$ the width of the log-normal size distribution.

\subsection{Calorimetry}

\begin{figure}[H]
\subfloat[ \label{graph//DSC-gesamt}]{ \includegraphics[width=0.9\linewidth]{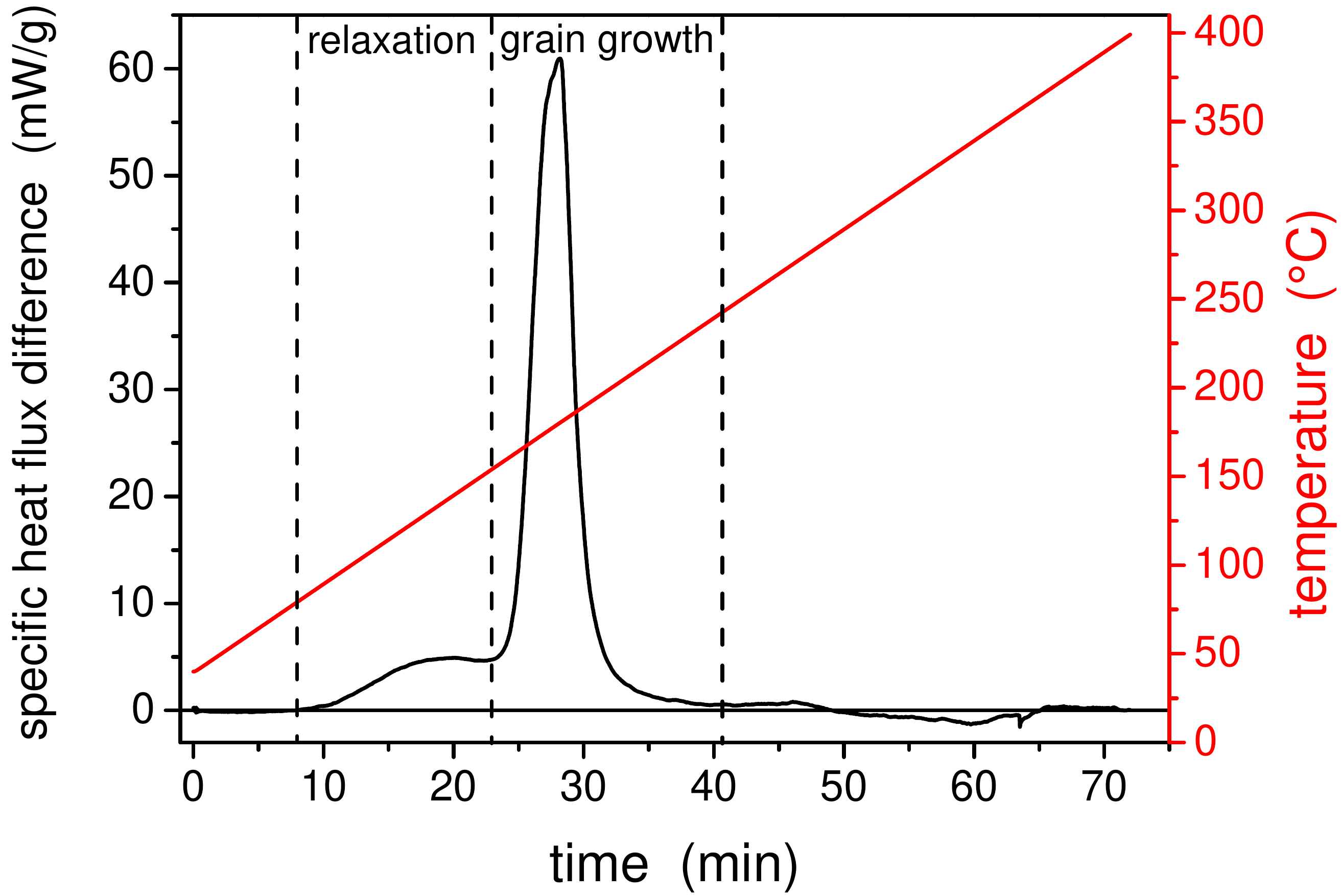}}
\hfill %
\subfloat[\label{graph//DSC-relaxation}]{\includegraphics[width=0.9\linewidth]{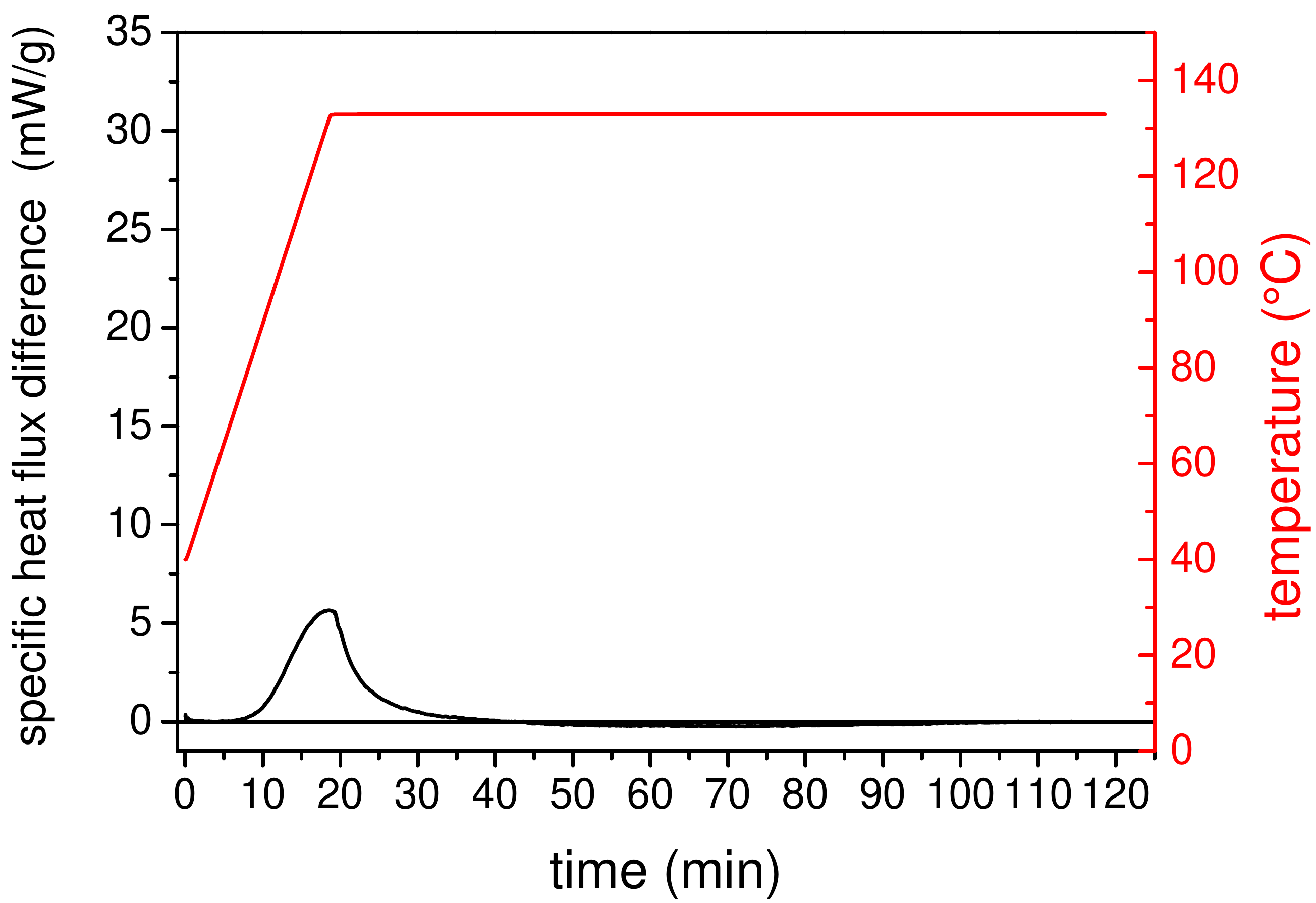}}
\hfill %
\subfloat[\label{graph//DSC-kornwachstum}]{\includegraphics[width=0.9\linewidth]{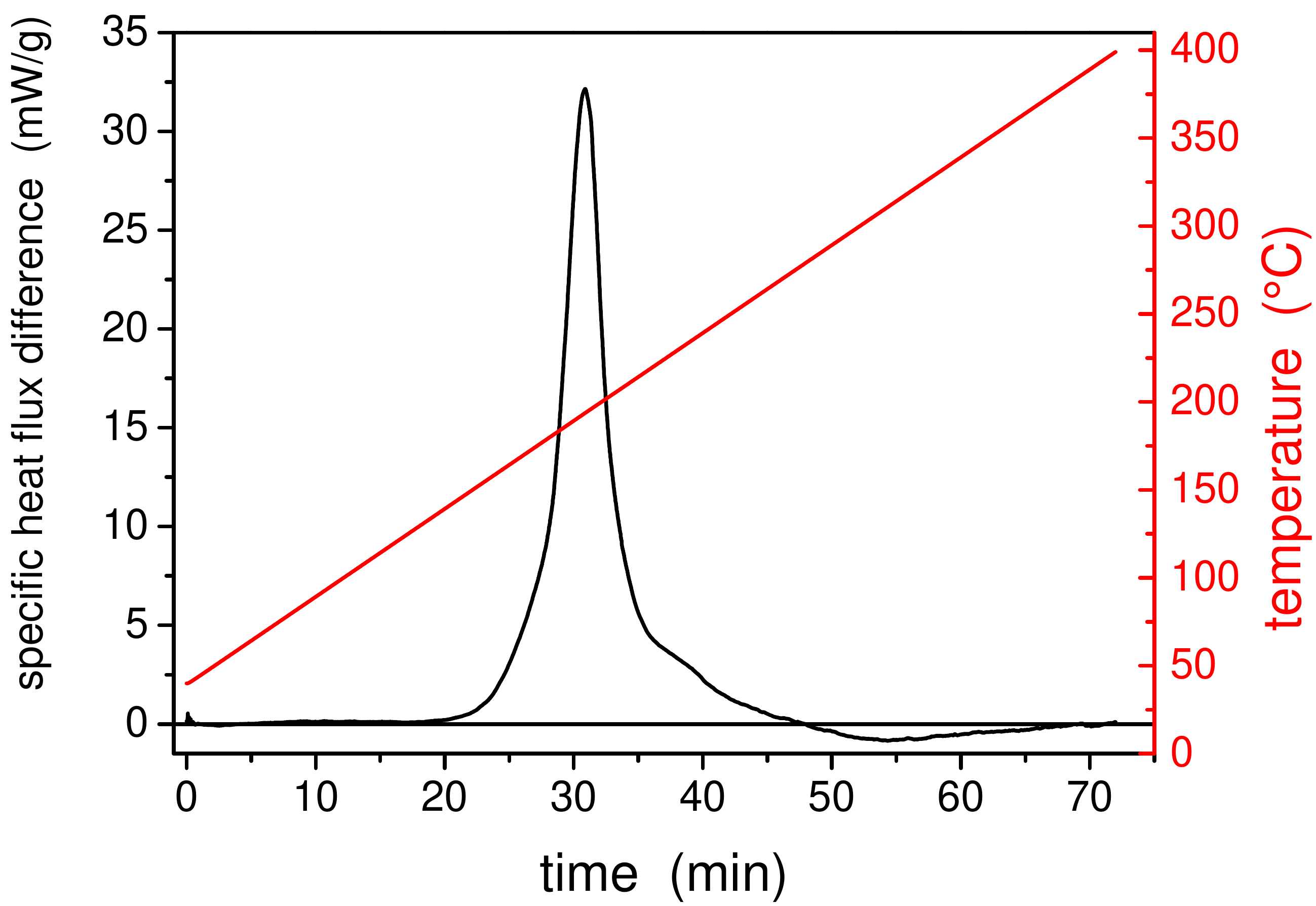}}
\hfill %
\caption[]{(a) DSC scan of an as prepared sample in the temperature range from $40^{\circ}C$ to $400^{\circ}C$ using a heating rate of $\unit[5]{^{\circ}C}/\mathrm{min}$. The first peak is due to structural relaxation without grain growth and the second is associated with grain growth. (b) Scanning a different sample from the same master specimen from $40^{\circ}C$ to $130^{\circ}C$ reveals a relaxation process but no activation of grain growth. (c) Running a scan of the relaxed sample (b) by using the scan-parameters specified in (a) generates a curvature driven grain-growth peak only.  
}  
\label{graph//DSC}
\end{figure}

Deformed samples were analyzed with respect to stored specific enthalpy by calorimetry using a DSC Q2000 differential scanning calorimeter from TA Instruments. We performed two scans following each other immediately by using the same time-temperature protocol consisting of a temperature ramp from $\unit[40]{^{\circ}C}$ to $\unit[400]{^{\circ}C}$ with a heating rate of $\unit[5]{^{\circ}C}/\mathrm{min}$. The signal of the second measurement then served as a baseline and was subtracted from the first measurement to exclusively obtain the signature of the evolution of irreversible processes. Enthalpy changes were obtained by integrating the heat flux difference over time. Upon heating, calorimetry reveals a thermally activated two-stage exothermic reaction for all as-prepared NC $\mathrm {Pd_{90}Au_{10}}$ samples (Fig. \ref{graph//DSC-gesamt}). The second stage is characterized by a much larger enthalpy release compared to the first-stage which appears at lower temperatures but is significantly broader. The enthalpy associated with the first stage is related to the relaxation of non-equilibrium GBs prevailing in the as-prepared state \cite{Tschoepe1993}. The signal of the second stage can be assigned to curvature-driven grain growth. Fortunately, both reactions can be separated thus enabling the analysis of the energetics of the related processes. \\
By carefully heating the sample well below the onset temperature of grain growth and annealing at this temperature $(\unit[130]{^{\circ}C}) $, we obtain the heat flux signal shown in Fig. \ref{graph//DSC-relaxation}. XRD measurements prove that after completion of this reaction the grain size remains basically unchanged. Thus, the measured enthalpy release is reflecting structural changes of the grain boundary core structure which has been denoted grain boundary relaxation \cite{Tschoepe1993}. Taking a sample after completion of the relaxation and heating it from room temperature to $\unit[400]{^{\circ}C}$, we observe the heat flux signal displayed in Fig. \ref{graph//DSC-kornwachstum}. This reaction is associated with massive grain growth up to several hundreds of micrometers.


\bibliographystyle{plain}                                                                   
\bibliography{LIT}

\end{multicols} 
 
\end{document}